\documentclass[prd,preprintnumbers,twocolumn,amsmath,nofootinbib,amssymb]{revtex4}
\usepackage{graphicx,color,dcolumn,booktabs,bm}
\usepackage{longtable,lscape}
\usepackage{txfonts}
\usepackage{overpic}
\usepackage{amssymb}
\usepackage{epstopdf}
\usepackage{indentfirst}
\usepackage{feynmf}   
\usepackage{slashed}  
\usepackage{cases}
\usepackage{color}
\usepackage{float}
\usepackage{multirow}
\usepackage{ulem}
\usepackage{graphicx,color,dcolumn,booktabs,bm}
\usepackage{epsfig,dsfont,amssymb,amsmath,amsfonts,amsbsy,mathrsfs}
\usepackage{pifont}

\graphicspath{{Figures/}} %

\usepackage{hyperref}
\hypersetup{colorlinks,citecolor=blue,anchorcolor=red,menucolor=red, linkcolor=red,filecolor=red,runcolor=red,urlcolor=blue,frenchlinks=red}

\makeatletter
\@addtoreset{equation}{section}
\makeatother

\allowdisplaybreaks

\begin{document}

\title{Electromagnetic characteristics as probes into the inner structures of\\the predicted $\Xi_c^{(',*)}D^{(*)}_s$ molecular states}

\author{Sheng-He Zhu$^{1,2,4}$}
\email{zhushh2024@lzu.edu.cn}
\author{Fu-Lai Wang$^{1,2,3,4}$}
\email{wangfulai@lzu.edu.cn}
\author{Xiang Liu$^{1,2,3,4}$\footnote{Corresponding author}}
\email{xiangliu@lzu.edu.cn}
\affiliation{$^1$School of Physical Science and Technology, Lanzhou University, Lanzhou 730000, China\\
$^2$Lanzhou Center for Theoretical Physics, Key Laboratory of Theoretical Physics of Gansu Province, Key Laboratory of Quantum Theory and Applications of MoE, Gansu Provincial Research Center for Basic Disciplines of Quantum Physics, Lanzhou University,
Lanzhou 730000, China\\
$^3$MoE Frontiers Science Center for Rare Isotopes, Lanzhou University, Lanzhou 730000, China\\
$^4$Research Center for Hadron and CSR Physics, Lanzhou University and Institute of Modern Physics of CAS, Lanzhou 730000, China}

\begin{abstract}
In this work, we conduct a systematic investigation of the electromagnetic properties, specifically the magnetic moments and the M1 radiative decay behavior, of the predicted $\Xi_c^{(',*)}D^{(*)}_s$-type double-charm hidden-strangeness molecular pentaquarks. The study is carried out within the framework of the constituent quark model to evaluate these electromagnetic observables, and our analysis incorporates three distinct scenarios: single-channel analysis, $S$-$D$ wave mixing analysis, and coupled-channel analysis. The calculated magnetic moments reveal characteristic patterns that reflect their underlying constituent configurations and provide sensitive probes for their quantum number assignments. Furthermore, we identify several M1 radiative decay channels with sizable widths that may offer promising signatures for future experimental detection. These M1 transitions also act as sensitive probes into their inner structures, displaying distinctive features that help differentiate between their constituent configurations and quantum number assignments. We anticipate that this study will stimulate experimental interest in exploring the electromagnetic properties of the $\Xi_c^{(',*)}D^{(*)}_s$ molecular states, thereby advancing our structural understanding of these exotic hadronic states.
\end{abstract}

\maketitle

\section{Introduction}

The quark model, formulated in the latter half of the 20th century, has long served as a highly successful framework for classifying the conventional hadrons. Specifically, the mesons as quark-antiquark pairs and the baryons as three-quark states. For decades, this model provided a systematic and comprehensive description of the observed hadronic spectrum. However, the advent of the 21st century brought a series of landmark experimental discoveries that fundamentally challenged this conventional hadron picture \cite{Karliner:2015ina,Meng:2022ozq,Liu:2013waa,Chen:2016qju,Hosaka:2016pey,Richard:2016eis,Chen:2022asf,Lebed:2016hpi,Olsen:2017bmm,Guo:2017jvc,Liu:2019zoy,Brambilla:2019esw,Liu:2024uxn,Wang:2025sic}. Several newly observed hadronic states, collectively termed the exotic hadrons, exhibit properties that cannot be explained within the simple quark-antiquark or three-quark configurations. Their emergence not only offers critical insights into the non-perturbative regime of strong interaction, but also opens new avenues for investigating the hadronic structures beyond the conventional hadron configurations.

The landscape of the exotic hadrons expanded significantly following the first observation of the charmonium-like state $X(3872)$ in 2003 \cite{Choi:2003ue}, with numerous exotic hadrons subsequently reported across various experimental facilities \cite{Karliner:2015ina,Meng:2022ozq,Liu:2013waa,Chen:2016qju,Hosaka:2016pey,Richard:2016eis,Chen:2022asf,Lebed:2016hpi,Olsen:2017bmm,Guo:2017jvc,Liu:2019zoy,Brambilla:2019esw,Liu:2024uxn,Wang:2025sic}. Remarkably, a substantial fraction of these observed hadronic states appears near two-hadron mass thresholds, suggesting their potential interpretation as the hadronic molecules. Particularly compelling evidence emerged in 2019 when the LHCb Collaboration reported three hidden-charm pentaquark states $P^N_{\psi}(4312)$, $P^N_{\psi}(4440)$, and $P^N_{\psi}(4457)$ \cite{Aaij:2019vzc}, which provides strong support for the hidden-charm molecular pentaquark hypothesis that had been previously predicted theoretically within the $\Sigma_c\bar{D}^{(*)}$ molecular picture \cite{Wu:2010jy,Wang:2011rga,Yang:2011wz,Wu:2012md}. Further expanding this family, LHCb subsequently announced the single-strangeness hidden-charm pentaquark states $P^\Lambda_{\psi s}(4459)$ \cite{LHCb:2020jpq} and $P^\Lambda_{\psi s}(4338)$ existing in the $J/\psi\Lambda$ invariant mass spectrum \cite{LHCb:2022ogu}, interpretable as the $\Xi_c\bar{D}^*$ and $\Xi_c\bar{D}$ molecular candidates, respectively \cite{Wu:2010vk,Weng:2019ynv,Wang:2019nvm,Azizi:2023foj,Wang:2022neq,Wang:2022mxy,Karliner:2022erb,Yan:2022wuz,Meng:2022wgl,Zhu:2022wpi,Chen:2022wkh,Ortega:2022uyu,Xiao:2019gjd,Anisovich:2015zqa,Chen:2016ryt,Hofmann:2005sw,Feijoo:2022rxf,Garcilazo:2022edi,Yang:2022ezl,Giachino:2022pws,Nakamura:2022jpd,Xiao:2022csb,Wang:2022gfb,Clymton:2022qlr,Chen:2022onm,Chen:2021spf,Ferretti:2021zis,Du:2021bgb,Chen:2021cfl,Hu:2021nvs,Lu:2021irg,Zou:2021sha,Wang:2021itn,Xiao:2021rgp,Zhu:2021lhd,Dong:2021juy,Wang:2020eep,Peng:2020hql,Chen:2020uif,Peng:2019wys,Chen:2020kco}. In addition, the 2022 discovery of the double-charm tetraquark state $T_{cc}(3875)^+$ \cite{LHCb:2021auc,LHCb:2021vvq}, explained as a $DD^*$ molecular candidate \cite{Manohar:1992nd,Ericson:1993wy,Tornqvist:1993ng,Janc:2004qn,Ding:2009vj,Molina:2010tx,Ding:2020dio,Li:2012ss,Xu:2017tsr,Liu:2019stu,Ohkoda:2012hv,Tang:2019nwv,Li:2021zbw,Chen:2021vhg,Ren:2021dsi,Xin:2021wcr,Chen:2021tnn,Albaladejo:2021vln,Dong:2021bvy,Baru:2021ldu,Du:2021zzh,Kamiya:2022thy,Padmanath:2022cvl,Agaev:2022ast,Ke:2021rxd,Zhao:2021cvg,Deng:2021gnb,Santowsky:2021bhy,Dai:2021vgf,Feijoo:2021ppq,Wang:2023ovj,Peng:2023lfw,Dai:2023cyo,Du:2023hlu,Kinugawa:2023fbf,Lyu:2023xro,Wang:2022jop,Wu:2022gie,Ortega:2022efc,Praszalowicz:2022sqx,Chen:2022vpo,Lin:2022wmj,Cheng:2022qcm,Mikhasenko:2022rrl}, further enriched this picture. The experimental observations of the hidden-charm pentaquark states and the double-charm tetraquark state have provided compelling evidence for the existence of the hadronic molecular states.

In previous theoretical studies, the experimental observation of the hidden-charm molecular pentaquark candidates alongside the double-charm molecular tetraquark candidate has motivated extensive investigations into the mass spectra of various double-charm molecular pentaquark states \cite{Wang:2023mdj,Wang:2022aga,Dong:2021bvy,Chen:2021kad,Yalikun:2023waw,Wang:2023ael,Wang:2023aob,Sheng:2024hkf,Wang:2025hhx,Wang:2023eng,Xu:2025mhc,Yang:2024okq,Wang:2024brl,Duan:2024uuf,Liu:2023clr,Shen:2022zvd}. Among them, a systematic analysis of the mass spectra of the $\Xi_c^{(',*)}D^{(*)}_s$-type double-charm hidden-strangeness molecular pentaquarks was conducted within the framework of the one-boson-exchange (OBE) model in Ref. \cite{Yalikun:2023waw}. Their results indicate the presence of several loosely bound states with $J^P={1/2}^-$, ${3/2}^-$, and ${5/2}^-$, which can be regarded as promising double-charm hidden-strangeness molecular pentaquark candidates.

Having established several $\Xi_c^{(',*)}D_s^{(*)}$ systems as promising molecular candidates \cite{Yalikun:2023waw}, a systematic investigation of their electromagnetic properties becomes essential. This natural progression from the mass spectra to the electromagnetic characteristics marks a critical step toward a comprehensive understanding of such exotic hadronic states. Electromagnetic properties of the hadrons offer uniquely sensitive probes into their inner structures, complementing information derived from their mass spectra, and playing a decisive role in determining their fundamental quantum numbers, such as spin-parity and isospin, as well as in identifying their constituent configurations. It is important to emphasize that while the study of the mass spectra determines whether the molecular formation is allowed, the electromagnetic observables probe the detailed structures of the resulting molecular systems. In this context, previous theoretical studies have extensively explored the electromagnetic properties of the heavy-flavor hadronic molecular states  \cite{Lai:2024jfe,Zhou:2022gra,Ozdem:2023htj,Ozdem:2022kei,Wang:2023aob,Wang:2023ael,Zhang:2025ame,Wang:2016dzu,Xu:2020flp,Wang:2022tib,Wang:2022nqs,Li:2021ryu,Gao:2021hmv,Wang:2023bek,Guo:2023fih,Li:2024wxr,Li:2024jlq,Sheng:2024hkf,Lei:2023ttd,Wang:2024sbw,Ozdem:2025hmb,Ozdem:2024ydl,Ozdem:2024yel,Ozdem:2024jty,Ozdem:2023okg,Ozdem:2023eyz,Ozdem:2022eds,Ozdem:2022ylm,Ozdem:2022yhi,Ozdem:2021vry,Ozdem:2021hka,Ozdem:2021yvo,Ozdem:2021ugy,Ozdem:2021btf,Ozdem:2018qeh,Mutuk:2024ltc,Mutuk:2024jxf,Mutuk:2024elj,Lei:2024geu,Liu:2003ab,Huang:2004tn,Zhu:2004xa,Ozdem:2025olj}, laying a solid foundation for further research.

In this work, we conduct a systematic investigation of the electromagnetic properties of the predicted $\Xi_c^{(',*)}D^{(*)}_s$-type double-charm hidden-strangeness molecular pentaquarks. Our study focuses on two fundamental electromagnetic observables: the magnetic moments and the M1 radiative decay behavior. Among them, the magnetic moments exhibit characteristic patterns that can distinguish between different constituent configurations and quantum number assignments. Meanwhile, the M1 radiative decays provide access to the transition matrix elements that are directly sensitive to the overlap of the wave functions including the color, flavor, spin-orbit, and spatial components corresponding to its internal degrees of freedom, thereby revealing key structural features of the hadrons. Methodologically, we employ the constituent quark model framework to calculate both the magnetic moments and the M1 radiative decay widths of the predicted $\Xi_c^{(',*)}D^{(*)}_s$ molecular pentaquarks, implementing a comprehensive approach that includes single-channel calculations, $S$-$D$ wave mixing effects, and coupled-channel contributions, which allows us to assess the robustness of our predictions. These investigations are expected to provide valuable insights for future experimental identification of these exotic hadron states.

The present paper is structured as follows. In Section \ref{section:2}, we systematically evaluate the magnetic moments of the predicted $\Xi_c^{(',*)}D^{(*)}_s$ molecular states within the constituent quark model framework, employing three distinct approaches: single-channel analysis, $S$-$D$ wave mixing analysis, and coupled-channel analysis. Section \ref{section:3} extends this methodology to examine the M1 radiative decay behavior of these predicted molecular states, maintaining the same theoretical framework and analytical hierarchy. Finally, Section \ref{section:4} provides a comprehensive summary of our principal findings.

\section{The magnetic moment properties of the predicted $\Xi_c^{(',*)}D^{(*)}_s$ molecules}\label{section:2}

In this work, we focus on the electromagnetic properties of the $\Xi_c^{(',*)}D^{(*)}_s$ molecular states. However, the investigation of the electromagnetic observables in the hadronic molecular states relies fundamentally on prior knowledge of their mass spectra. In particular, the spatial wave functions of the hadronic molecular states serve as crucial inputs for discussing their electromagnetic characteristics. For this reason, we first revisit the existing theoretical results on the mass spectra of the $\Xi_c^{(',*)}D^{(*)}_s$ molecular states, drawing primarily on the systematic study presented in Ref. \cite{Yalikun:2023waw}, which serves as the foundational input for our subsequent electromagnetic analysis.

In Ref. \cite{Yalikun:2023waw}, the mass spectra of the $\Xi_c^{(',*)}D^{(*)}_s$-type double-charm hidden-strangeness molecular pentaquark candidates with $J^P={1/2}^-$, ${3/2}^-$, and ${5/2}^-$ were systematically investigated within the OBE model, incorporating exchanges of the light mesons $\sigma$, $\eta$, and $\phi$. As reported in previous work \cite{Yalikun:2023waw}, the ten near-threshold loosely bound states $\Xi_cD_s({1/2}^-)$, $\Xi_c^{'}D_s({1/2}^-)$, $\Xi_cD^{*}_s({1/2}^-)$, $\Xi_cD^{*}_s({3/2}^-)$, $\Xi_c^*D_s({3/2}^-)$, $\Xi_c^{'}D^{*}_s({1/2}^-)$, $\Xi_c^{'}D^{*}_s({3/2}^-)$, $\Xi_c^*D^{*}_s({1/2}^-)$, $\Xi_c^*D^{*}_s({3/2}^-)$, and $\Xi_c^*D^{*}_s({5/2}^-)$ were identified in single-channel analysis with $S$-$D$ wave mixing effects, using a phenomenologically reasonable cutoff parameter from 1.0 to 2.5 GeV. A subsequent coupled-channel analysis further corroborated the viability of several $\Xi_c^{(',*)}D^{(*)}_s$ states as promising double-charm hidden-strangeness molecular pentaquark candidates. Given the consistency between our calculated mass spectra for the $\Xi_c^{(',*)}D^{(*)}_s$-type double-charm hidden-strangeness molecular pentaquarks and those reported in Ref. \cite{Yalikun:2023waw}, we do not reproduce the detailed mass spectrum results in this work.

\begin{figure}[hbtp]
\centering
\includegraphics[width=8.6cm]{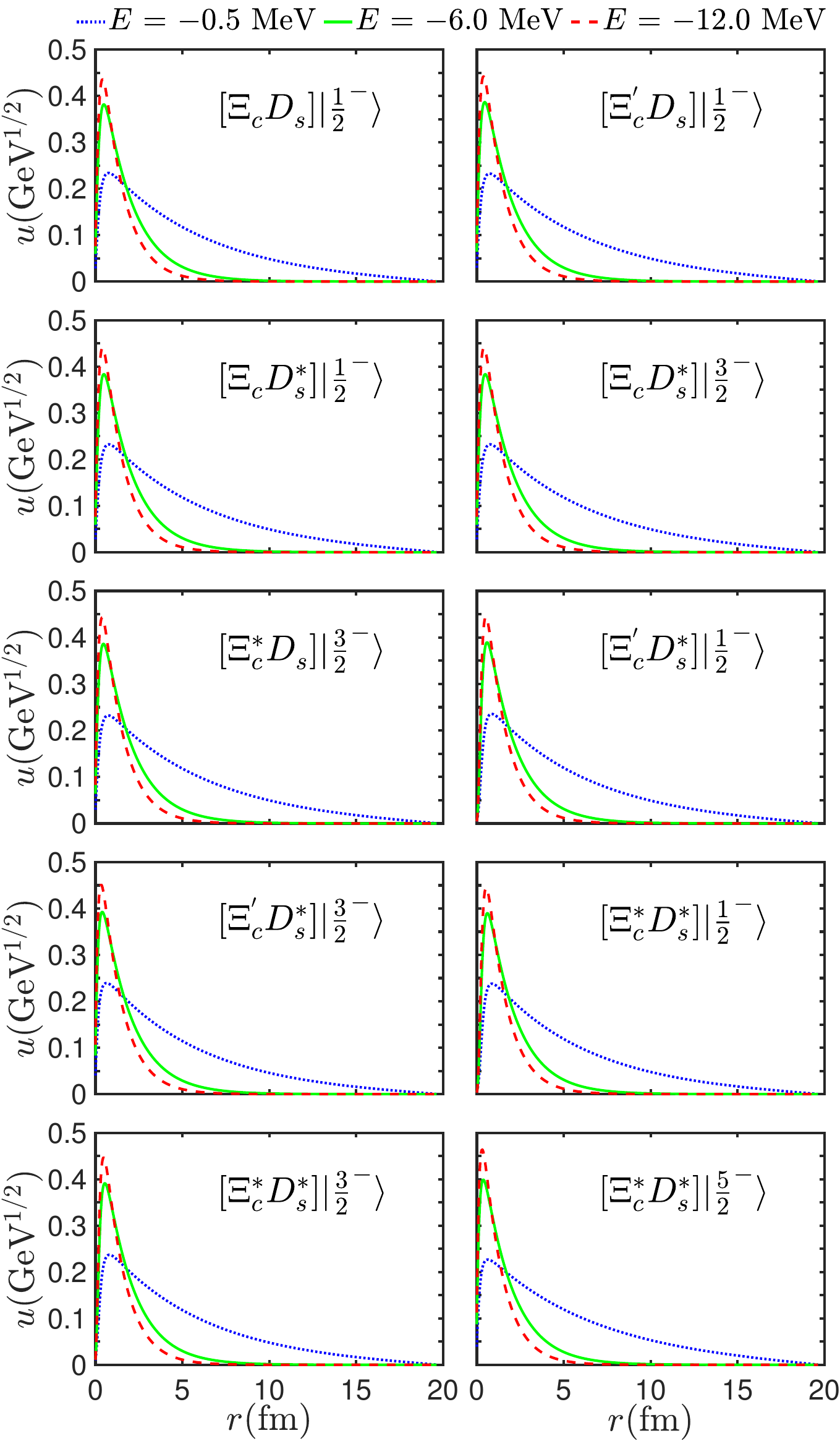}
\caption{The spatial wave functions $u(r)$ of ten predicted $\Xi_c^{(',*)}D^{(*)}_s$ molecular pentaquarks obtained from the single-channel analysis at three representative binding energies of $-0.5$, $-6.0$, and $-12.0$ MeV.}
\label{fig:singlefun}
\end{figure}

{However, the binding energies of these predicted $\Xi_c^{(',*)}D^{(*)}_s$ molecular states remain ambiguous, due to the lack of experimental observation of such states. Within the constituent quark model, the determination of the electromagnetic properties of the $\Xi_c^{(',*)}D^{(*)}_s$ molecular states relies on the spatial wave functions and the masses of these molecular states. Notably, both the spatial wave functions and the masses are closely related to their binding energies \cite{Chen:2017xat}. Although the exact binding energies of the $\Xi_c^{(',*)}D^{(*)}_s$ molecular states are unknown, it is theoretically well established that such molecular states are loosely bound \cite{Chen:2016qju}, characterized by small binding energies and spatially extended wave functions. In this study, we consider three representative binding energies $-0.5$, $-6.0$, and $-12.0$ MeV to investigate the electromagnetic properties of the $\Xi_c^{(',*)}D^{(*)}_s$ molecular states.} In the following, we discuss their spatial wave functions. As an illustration, Fig. \ref{fig:singlefun} shows the spatial wave functions $u(r)$ of ten predicted $\Xi_c^{(',*)}D^{(*)}_s$ molecular pentaquarks corresponding to these three binding energies, {which can be obtained from the quantitative study of their mass spectrum.} As can be observed from Fig. \ref{fig:singlefun}, a smaller binding energy of the $\Xi_c^{(',*)}D^{(*)}_s$ molecular pentaquarks corresponds to a more spatially extended wave function. This behavior is consistent with the theoretical picture of the hadronic molecular states, where the weakly bound states are expected to exhibit broader spatial distributions \cite{Chen:2016qju}.

In alignment with the approaches for the mass spectrum calculations of the $\Xi_c^{(',*)}D^{(*)}_s$ molecular pentaquarks in Ref. \cite{Yalikun:2023waw} and in pursuit of a comprehensive analysis of their electromagnetic properties, we incorporate both $S$-$D$ wave mixing effects and coupled-channel effects in our study. These physical mechanisms may play essential role for a rigorous treatment of the electromagnetic characteristics of such systems. To facilitate the subsequent discussion, we list the specific $S$- and $D$-wave channels, denoted as $|^{2S+1}L_J\rangle$, considered for the $\Xi_c^{(',*)}D^{(*)}_s$ molecular pentaquarks when $S$-$D$ wave mixing effects are included:
\begin{align*}
	    &\Xi_c^{(')}D_s:&&\hspace{-1em} J^P=\frac{1}{2}^-:|^2S_{\frac{1}{2}}\rangle. \\
	  	&\Xi^*_cD_s:&&\hspace{-1em}J^P=\frac{3}{2}^-:|^4S_{\frac{3}{2}}\rangle,|^4D_{\frac{3}{2}}\rangle. \\
	  	&\Xi_c^{(')}D^{*}_s: &&\hspace{-1em}J^P=\frac{1}{2}^-:|^2S_{\frac{1}{2}}\rangle,|^4D_{\frac{1}{2}}\rangle,\\
	  	&&&\hspace{-1em}J^P=\frac{3}{2}^-:|^4S_{\frac{3}{2}}\rangle,|^2D_{\frac{3}{2}}\rangle,|^4D_{\frac{3}{2}}\rangle.\\
	  	&\Xi_c^{*}D^{*}_s:&&\hspace{-1em}J^P=\frac{1}{2}^-:|^2S_{\frac{1}{2}}\rangle,|^4D_{\frac{1}{2}}\rangle,|^6D_{\frac{1}{2}}\rangle,\\
	  	&&&\hspace{-1em}J^P=\frac{3}{2}^-:|^4S_{\frac{3}{2}}\rangle,|^2D_{\frac{3}{2}}\rangle,|^4D_{\frac{3}{2}}\rangle,|^6D_{\frac{3}{2}}\rangle,\\
	  	&&&\hspace{-1em}J^P=\frac{5}{2}^-:|^6S_{\frac{5}{2}}\rangle,|^2D_{\frac{5}{2}}\rangle,|^4D_{\frac{5}{2}}\rangle,|^6D_{\frac{5}{2}}\rangle.
\end{align*}
Here, the notation $|^{2S+1}L_J\rangle$ labels the channels in the discussed system, where $S$, $L$, and $J$ represent the spin, orbital, and total angular momentum quantum numbers, respectively. The spin-orbital wave function for a channel $|^{2S+1}L_J\rangle$ is constructed as
\begin{equation}
|^{2S+1}L_J\rangle=\sum\limits_{S_z,L_z}C^{J,J_z}_{S,S_z;L,L_z}|S,S_z\rangle|L,L_z\rangle,\end{equation}
where $C^{J,J_z}_{S,S_z;L,L_z}$ is the Clebsch-Gordan coefficient, $|S,S_z\rangle$ is the spin wave function, and $|L,L_z\rangle$ is the orbital wave function.

\subsection{Framework for calculating the hadronic magnetic moment in the constituent quark model}

The study of the hadronic magnetic moments has garnered substantial theoretical interest, with investigations employing various approaches, such as the lattice QCD simulations, the chiral perturbation theory,  the constituent quark model, the QCD sum rules, the Bag model, and so on \cite{Meng:2022ozq}. Among these, the constituent quark model has been successful in quantitatively reproducing the experimentally measured magnetic moments of the octet and decuplet baryons \cite{Schlumpf:1993rm, Kumar:2005ei, Ramalho:2009gk}. Motivated by this success, we employ the constituent quark model in this work to systematically study the magnetic moments of the predicted $\Xi_c^{(',*)}D^{(*)}_s$ molecular pentaquarks. Below, we outline the formalism used to calculate the magnetic moments of the predicted $\Xi_c^{(',*)}D^{(*)}_s$ molecular candidates within the constituent quark model.

Within the constituent quark model, the magnetic moment of a hadronic molecule $[AB]$, denoted as $\mu_{[AB]}$, is given by the expectation value \cite{Liu:2003ab,Huang:2004tn,Zhu:2004xa,Wang:2016dzu,Li:2021ryu,Zhou:2022gra,Wang:2022tib,Gao:2021hmv,Wang:2023bek,Lai:2024jfe,Mutuk:2024ltc,Mutuk:2024elj,Li:2024wxr,Lei:2024geu,Lei:2023ttd,Zhang:2025ame,Wang:2022nqs,Mutuk:2024jxf,Guo:2023fih}:
\begin{equation}
	\mu_{[AB]} = \left\langle J_{[AB]},\,J_{[AB]} \left| \sum\limits_{j}\hat{\mu}_{zj}^{\text{spin}}+\hat{\mu}_z^{\text{orbital}} \right| J_{[AB]},\,J_{[AB]} \right\rangle.
\end{equation}
Here, $|J_{[AB]},\,J_{[AB]}\rangle$ denotes the total wave function of the discussed molecular state $[AB]$, constructed as the direct product of the color, flavor, spin-orbit, and spatial components corresponding to its internal degrees of freedom. The $z$-component for both the spin magnetic moment operator $\hat{\mu}_{zj}^{\text{spin}}$ and the orbital magnetic moment operator $\hat{\mu}_z^{\text{orbital}}$ are defined as \cite{Liu:2003ab,Huang:2004tn,Zhu:2004xa,Wang:2016dzu,Li:2021ryu,Zhou:2022gra,Wang:2022tib,Gao:2021hmv,Wang:2023bek,Lai:2024jfe,Mutuk:2024ltc,Mutuk:2024elj,Li:2024wxr,Lei:2024geu,Lei:2023ttd,Zhang:2025ame,Wang:2022nqs,Mutuk:2024jxf,Guo:2023fih}:
\begin{eqnarray}
    \hat{\mu}_{zj}^{\text{spin}} &=& \frac{e_j}{2m_j}\hat{\sigma}_{zj},\label{eq:muspin}\\
    \hat{\mu}_z^{\text{orbital}} &=& \left(\frac{m_{A}}{m_{A} + m_{B}}\frac{e_{B}}{2m_{B}} + \frac{m_{B}}{m_{A} + m_{B}}\frac{e_{A}}{2m_{A}}\right)\hat{L}_z.\label{eq:muorbital}
\end{eqnarray}
The charge $e_j$, the mass $m_j$, and the $z$-component of the Pauli spin operator $\hat{\sigma}_{zj}$ are assigned to the $j$-th quark constituent of the hadron. The $z$-component of the relative orbital angular momentum between the constituent hadrons $A$ and $B$ is denoted by $\hat{L}_z$.

As defined in Eq.~(\ref{eq:muspin}),  the constituent quark masses are essential inputs for calculating the hadronic magnetic moments within the constituent quark model. In this work, we adopt the following set of constituent quark masses: $m_u=336.00$ MeV, $m_d=336.00$ MeV, $m_s=450.00$ MeV, and $m_c=1680.00$ MeV \cite{Kumar:2005ei}. This parameter set has not only successfully reproduced the magnetic moments of the octet and decuplet baryons where experimental data exist \cite{Schlumpf:1993rm, Kumar:2005ei, Ramalho:2009gk}, but has also been widely employed in recent systematic studies of the magnetic moments for the hadronic molecular states \cite{Sheng:2024hkf,Li:2021ryu,Wang:2023bek,Gao:2021hmv,Wang:2022tib,Wang:2023aob,Wang:2023ael,Wang:2022nqs,Zhou:2022gra,Lai:2024jfe,Zhang:2025ame,Wang:2024sbw}. When accounting for $S$-$D$ wave mixing effects, the masses of the constituent hadrons within the molecular state also become crucial, as indicated in Eq.~(\ref{eq:muorbital}). In our numerical analysis, we use $m_{\Xi_c^+}=2467.71$ MeV, $m_{\Xi_c^0}=2470.44$ MeV, $m_{\Xi_c^{\prime+}}=2578.20$ MeV, $m_{\Xi_c^{\prime0}}=2578.70$ MeV, $m_{\Xi_c^{*+}}=2645.10$ MeV, $m_{\Xi_c^{*0}}=2646.16$ MeV, $m_{D_s^+}=1968.35$ MeV, and $m_{D_s^{*+}}=2112.20$ MeV for the baryons and mesons masses, taking the values from the Particle Data Group \cite{ParticleDataGroup:2024cfk}.

\subsection{Calculated magnetic moments of the predicted $\Xi_c^{(',*)}D^{(*)}_s$ molecular pentaquarks}

Building upon the theoretical framework outlined in the preceding subsection, we now present a systematic analysis of the magnetic moments for the $\Xi_c^{(',*)}D^{(*)}_s$ molecular pentaquarks. The corresponding numerical results are organized according to three progressively refined treatments: single-channel analysis, $S$-$D$ wave mixing analysis, and coupled-channel analysis.

\subsubsection{Single-channel analysis}

To calculating the magnetic moments of the $\Xi_c^{(',*)}D^{(*)}_s$ molecular pentaquarks, we begin by constructing their flavor and spin wave functions. It should be noted that the color wave function of a hadronic molecular state is taken as a color singlet. Furthermore, in single-channel analysis, the spatial wave function is required to satisfy the normalization condition. The flavor wave functions of the $\Xi_c^{(',*)}D^{(*)}_s$ systems, denoted as $|I, I_z \rangle$, are given by:
  \begin{equation*}
      \Xi_c^{(',*)+} D_s^{(*)+} : \left| \frac{1}{2}, \frac{1}{2} \right\rangle~~~~{\rm  and}~~~~
      \Xi_c^{(',*)0} D_s^{(*)+} :\left| \frac{1}{2}, -\frac{1}{2} \right\rangle,
\end{equation*}
where $I$ and $I_z$ represent the total isospin and its $z$-component, respectively. And the spin wave functions $| S, S_z \rangle$ for the $\Xi_c^{(',*)}D^{(*)}_s$ systems are constructed via the following coupling scheme:
 \begin{eqnarray*}
	\Xi_c^{(')} D_s&:& \left| \frac{1}{2}, S_{\Xi_c^{(')}} \right\rangle \left| 0, 0 \right\rangle, \\
    \Xi_c^{(')} D_s^{*}&:& \sum\limits_{S_{\Xi_c^{(')}},S_{D_s^{*}}} C^{S,S_z}_{\frac{1}{2}, S_{\Xi_c^{(')}}; 1,S_{D_s^{*}}} \left| \frac{1}{2}, S_{\Xi_c^{(')}} \right\rangle \left| 1, S_{D_s^{*}} \right\rangle,\\
	\Xi_c^{*} D_s&:& \left| \frac{3}{2}, S_{\Xi_c^{*}} \right\rangle \left| 0, 0 \right\rangle,\\
    \Xi_c^{*} D_s^{*}&:& \sum\limits_{S_{\Xi_c^{*}},S_{D_s^{*}}} C^{S,S_z}_{\frac{3}{2}, S_{\Xi_c^{*}}; 1,S_{D_s^{*}}} \left| \frac{3}{2}, S_{\Xi_c^{*}} \right\rangle \left| 1, S_{D_s^{*}} \right\rangle.
\end{eqnarray*}
Here, $S_A$ denotes the $z$-component of the spin for the corresponding hadron $A$, and $C^{S,S_z}_{s_1, m_1; s_2, m_2}$ represents the Clebsch-Gordan coefficient in the coupling scheme.

\begin{figure}[hbtp]
\centering
\includegraphics[width=8.6cm]{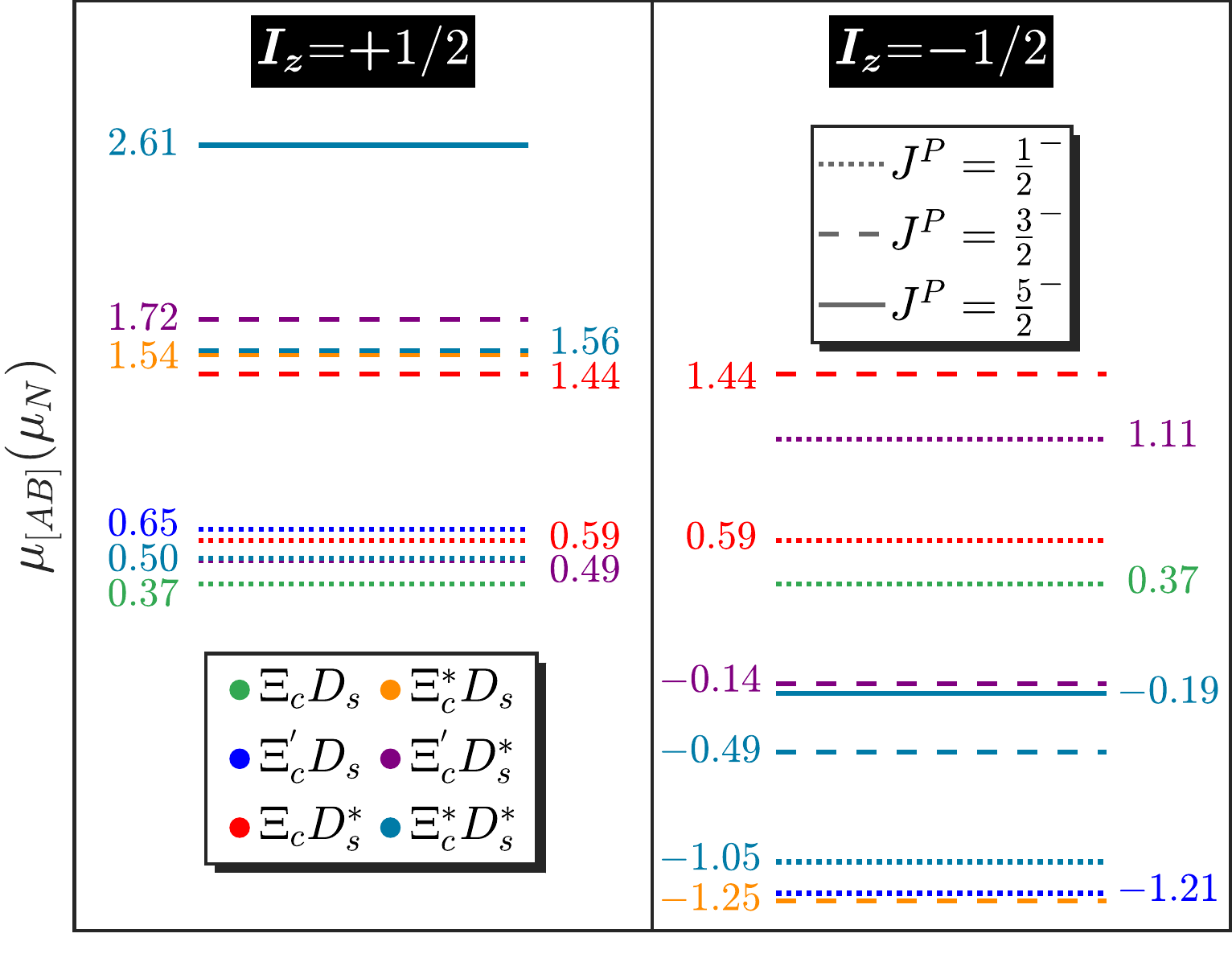}
\caption{The magnetic moments $\mu_{[AB]}$ of the predicted $\Xi_c^{(',*)}D^{(*)}_s$ molecular states in single-channel analysis.}
\label{fig:singlemu}
\end{figure}

In Fig. \ref{fig:singlemu}, we present the magnetic moments of the predicted $\Xi_c^{(',*)}D^{(*)}_s$ molecular states obtained in single-channel analysis. All values are expressed in units of $\mu_N=e/2m_p$ with $m_p=938.00$ MeV \cite{ParticleDataGroup:2024cfk}. Several noteworthy features emerge from the results:
\begin{itemize}
  \item In single-channel approximation,  the magnetic moments of the $\Xi_c^{(',*)}D^{(*)}_s$ molecular states are given by the sum of the magnetic moments of the baryons $\Xi_c^{(',*)}$ and the mesons $D^{(*)}_s$, i.e.,
\begin{eqnarray*}
\mu_{[\Xi_c^{(')}D_s]|\frac{1}{2}^-\rangle}&=&\mu_{\Xi^{(')}_c}+\mu_{D_s},\\
\mu_{[\Xi^*_cD_s]|\frac{3}{2}^-\rangle}&=&\mu_{\Xi^*_c}+\mu_{D_s},\\
\mu_{[\Xi_c^{(')}D^{*}_s]|\frac{1}{2}^-\rangle}&=&-\frac{1}{3}\mu_{\Xi_c^{(')}}+\frac{2}{3}\mu_{D_s^{*}},\\
\mu_{[\Xi_c^{(')}D^{*}_s]|\frac{3}{2}^-\rangle}&=&\mu_{\Xi_c^{(')}}+\mu_{D_s^{*}},\\
\mu_{[\Xi_c^{*}D^{*}_s]|\frac{1}{2}^-\rangle}&=&\frac{5}{9}\mu_{\Xi_c^{*}}-\frac{1}{3}\mu_{D_s^{*}},\\
\mu_{[\Xi_c^{*}D^{*}_s]|\frac{3}{2}^-\rangle}&=&\frac{11}{15}\mu_{\Xi_c^{*}}+\frac{2}{5}\mu_{D_s^{*}},\\
\mu_{[\Xi_c^{*}D^{*}_s]|\frac{5}{2}^-\rangle}&=&\mu_{\Xi_c^{*}}+\mu_{D_s^{*}}.
\end{eqnarray*}
Thus, each constituent hadron contributes its intrinsic magnetic moment to the total value of the discussed molecular state.
  \item The magnetic moments of the $\Xi_c D_s$ molecular state with $J^P = 1/2^-$ are identical for the isospin projections $I_z=+1/2$ and $I_z=-1/2$, a feature stemming from the equality of the constituent magnetic moments of $\Xi_c^+$ and $\Xi_c^0$. Furthermore, the $\Xi_cD_s^{*}$ molecular states with $J^P={1}/{2}^-$ and ${3}/{2}^-$ exhibit a similar characteristic under analogous conditions.
  \item  Notably, significant differences in the magnetic moments are observed between the hadronic molecular states that share identical constituent hadrons but differ in spin-parity quantum numbers, as exemplified by the $\Xi_c^{\prime}D_s^{*}$ molecular state with $J^P={1}/{2}^-$ versus the $\Xi_c^{\prime}D_s^{*}$ molecular state with $J^P={3}/{2}^-$, the $\Xi_c^{*}D_s^{*}$ molecular state with $J^P={1}/{2}^-$ versus the $\Xi_c^{*}D_s^{*}$ molecular state with $J^P={3}/{2}^-$, and so on. Similarly, distinct magnetic moments are found for the hadronic molecular states possessing the same spin-parity quantum numbers but composed of different constituent hadrons, such as the $\Xi_cD_s$ molecular state with $J^P={1}/{2}^-$ and the $\Xi_c^{\prime}D_s$ molecular state with $J^P={1}/{2}^-$, the $\Xi_c^{\prime}D_s^{*}$ molecular state with $J^P={3}/{2}^-$ and the $\Xi_c^{*}D_s^{*}$ molecular state with $J^P={3}/{2}^-$, and so on. These systematic variations demonstrate that the magnetic moment serves as a sensitive probe for discriminating both the spin-parity assignments and the constituent configurations of the $\Xi_c^{(',*)}D^{(*)}_s$ molecular pentaquarks.
\end{itemize}

{As important input parameters, the constituent quark masses play a crucial role in determining the electromagnetic properties of the predicted $\Xi_c^{(',*)}D^{(*)}_s$ molecular states within the constituent quark model. However, due to the lack of relevant experimental data, these masses cannot be determined with high precision. These parameter set adopted in this work has not only successfully reproduced the magnetic moments of the octet and decuplet baryons, for which experimental data are available, but has also been widely employed in recent systematic studies of the electromagnetic properties of the hadronic molecular states. To assess the impact of parametric uncertainties, we vary all constituent quark masses within a physically motivated range of approximately $\pm 10\%$ when discussing the magnetic moments of the $\Xi_c^{(',*)}D^{(*)}_s$ molecular states in single-channel analysis. This range covers the typical variations found in different constituent quark masses parametrization in Ref. \cite{Majethiya:2009vx}. The resulting ranges for the magnetic moments will be indicated as superscripts and subscripts attached to the central values (see Table \ref{table1}). From Table \ref{table1}, we find that while the constituent quark masses affect the magnetic moments of the $\Xi_c^{(',*)}D^{(*)}_s$ molecular states, their influence is not significant.}

\renewcommand\tabcolsep{0.80cm}
\renewcommand{\arraystretch}{1.50}
\begin{table}[!htbp]
\centering
\caption{The magnetic moments of the predicted $\Xi_c^{(',*)}D^{(*)}_s$ molecular states in single-channel analysis, in units of $\mu_N$. The central values are calculated using the fixed constituent quark masses, while the superscripts and subscripts denote the upper and lower bounds obtained from varying the constituent quark masses within approximately $\pm 10\%$.}\label{table1}
\begin{tabular}{ccc}
  \toprule[1pt]\toprule[1pt]
Molecules & $I_z=+\frac{1}{2}$ & $I_z=-\frac{1}{2}$ \\ [1pt] \hline
  $[\Xi_c^{} D_s^{} ]| \frac{1}{2}^- \rangle $ &$0.37^{+0.04}_{-0.03}$&$0.37^{+0.04}_{-0.03}$\\
  $[\Xi_c^{'} D_s^{} ]| \frac{1}{2}^- \rangle $ &$0.65^{+0.19}_{-0.18}$&$-1.21^{+0.11}_{-0.13}$\\
  $[\Xi_c^{} D_s^{*} ]| \frac{1}{2}^- \rangle $ &$0.59^{+0.09}_{-0.08}$&$0.59^{+0.09}_{-0.08}$\\
  $[\Xi_c^{} D_s^{*} ]| \frac{3}{2}^- \rangle $ &$1.44^{+0.16}_{-0.13}$&$1.44^{+0.16}_{-0.13}$\\
  $[\Xi_c^{*} D_s^{} ]| \frac{3}{2}^- \rangle $ &$1.54^{+0.31}_{-0.28}$&$-1.25^{+0.19}_{-0.21}$\\
   $[\Xi_c^{'} D_s^{*} ]| \frac{1}{2}^- \rangle $ &$0.49^{+0.14}_{-0.13}$&$1.11^{+0.12}_{-0.10}$\\
  $[\Xi_c^{'} D_s^{*} ]| \frac{3}{2}^- \rangle $ &$1.72^{+0.31}_{-0.28}$&$-0.14^{+0.23}_{-0.23}$\\
   $[\Xi_c^{*} D_s^{*} ]| \frac{1}{2}^- \rangle $ &$0.50^{+0.21}_{-0.20}$&$-1.05^{+0.14}_{-0.16}$\\
  $[\Xi_c^{*} D_s^{*} ]| \frac{3}{2}^- \rangle $ &$1.56^{+0.28}_{-0.24}$&$-0.49^{+0.19}_{-0.20}$\\
    $[\Xi_c^{*} D_s^{*} ]| \frac{5}{2}^- \rangle $ &$2.61^{+0.43}_{-0.38}$&$-0.19^{+0.31}_{-0.31}$\\
     \toprule[1pt]\toprule[1pt]
    \end{tabular}
\end{table}

\subsubsection{$S$-$D$ wave mixing analysis}

We now extend our analysis of the magnetic moments for the predicted $\Xi_c^{(',*)}D^{(*)}_s$ molecular states by incorporating contributions from the $D$-wave channels. It should be noted that the tensor force is absent in the interactions of the $\Xi_cD_s^{*}$ and $\Xi_c^{*}D_s$ systems \cite{Yalikun:2023waw}, the $D$-wave channel contribution is consequently zero in both of these discussed systems. Thus, we only consider the $\Xi_c^{'}D_s^{*}$ molecular states with $J^P=({1/2}^-,\,{3/2}^-)$ and $\Xi_c^{*}D_s^{*}$ molecular states with $J^P=({1/2}^-,\,{3/2}^-,\,{5/2}^-)$ in $S$-$D$ wave mixing case.

\begin{figure}[hbtp]
\centering
\includegraphics[width=8.6cm]{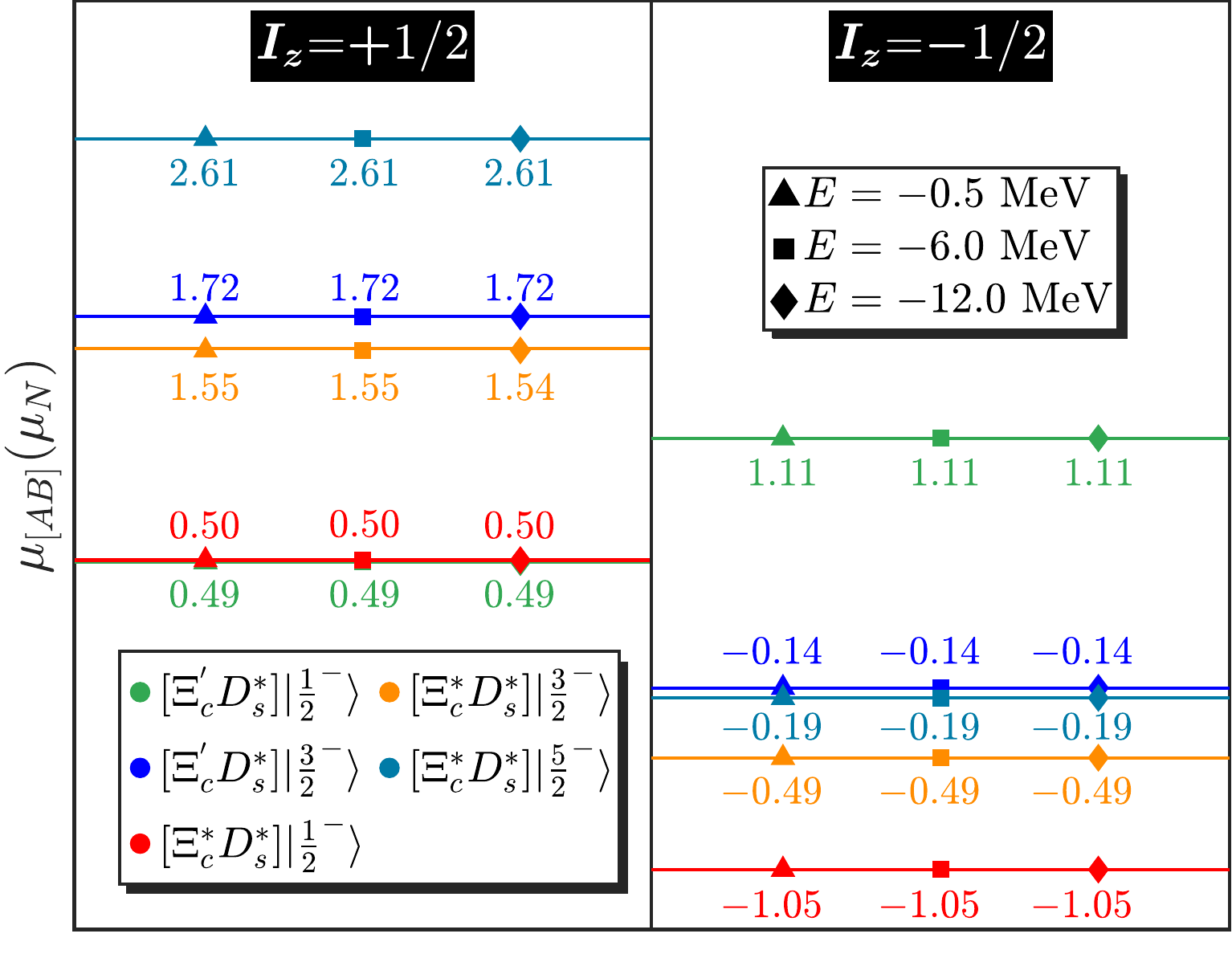}
\caption{The magnetic moments $\mu_{[AB]}$ of the predicted $\Xi_c^{(',*)}D^{(*)}_s$ molecular states in $S$-$D$ wave mixing analysis. The corresponding single-channel results are shown as the horizontal lines for comparison.}
\label{fig:sdmu}
\end{figure}

When studying the magnetic moments of the $\Xi_c^{(',*)}D^{(*)}_s$ molecular pentaquarks with $S$-$D$ wave mixing effects included, the spatial wave functions of both the $S$- and $D$-wave components are required as essential inputs. For this purpose, we adopt three typical binding energies of $-0.5$, $-6.0$, and $-12.0$ MeV to systematically examine their resulting magnetic moments under $S$-$D$ wave mixing analysis. The corresponding values for the $\Xi_c^{(',*)}D^{(*)}_s$ molecular pentaquarks are summarized in Fig. \ref{fig:sdmu}.

Through a systematic comparison of the magnetic moments of the
$\Xi_c^{(',*)}D^{(*)}_s$ molecular pentaquarks obtained from single-channel analysis and $S$-$D$ wave mixing analysis, we observe that the inclusion of $S$-$D$ wave mixing effects leads to only minor modifications. Moreover, the resulting magnetic moments show negligible dependence on the chosen binding energies. This behavior can be attributed to the very small $D$-wave component in the total spatial wave function, which leads to a correspondingly minor contribution to the overall magnetic moment.

\subsubsection{Coupled-channel analysis}

Having systematically analyzed the magnetic moment properties of the $\Xi_c^{(',*)}D^{(*)}_s$ molecular pentaquarks through both single-channel analysis and $S$-$D$ wave mixing analysis, we now proceed to include coupled-channel effects in our investigation.
For the $\Xi_cD_s$ state with $J^P={1/2}^-$ and the $\Xi_cD_s^{*}$ state with $J^P={3/2}^-$, when coupled-channel effects are incorporated into the analysis of their mass spectra, the resulting numerical outcomes are found to be physically inconsistent. This discrepancy arises primarily because the dominant channel in the configuration does not correspond to the one with the lowest threshold. Such a configuration leads to an anomalously small spatial size of the discussed systems, which contradicts the established picture of the hadronic molecular states as the spatially extended \cite{Chen:2016qju}. A detailed discussion of this issue can be found in Refs. \cite{Chen:2017xat,Wang:2021hql}. Consequently, in our subsequent study of the magnetic moments of the
$\Xi_c^{(',*)}D^{(*)}_s$ molecular pentaquarks incorporating coupled-channel effects, we focus on the $\Xi_c^{'}D_s$ molecular state with $J^P={1}/{2}^-$, the $\Xi_cD_s^{*}$ molecular state with $J^P={1}/{2}^-$, the $\Xi_c^{*}D_s$ molecular state with $J^P={3}/{2}^-$, the $\Xi_c^{'}D_s^{*}$ molecular state with $J^P={1}/{2}^-$, and the $\Xi_c^{'}D_s^{*}$ molecular state with $J^P={3}/{2}^-$.

\begin{figure}[hbtp]
\centering
\includegraphics[width=8.6cm]{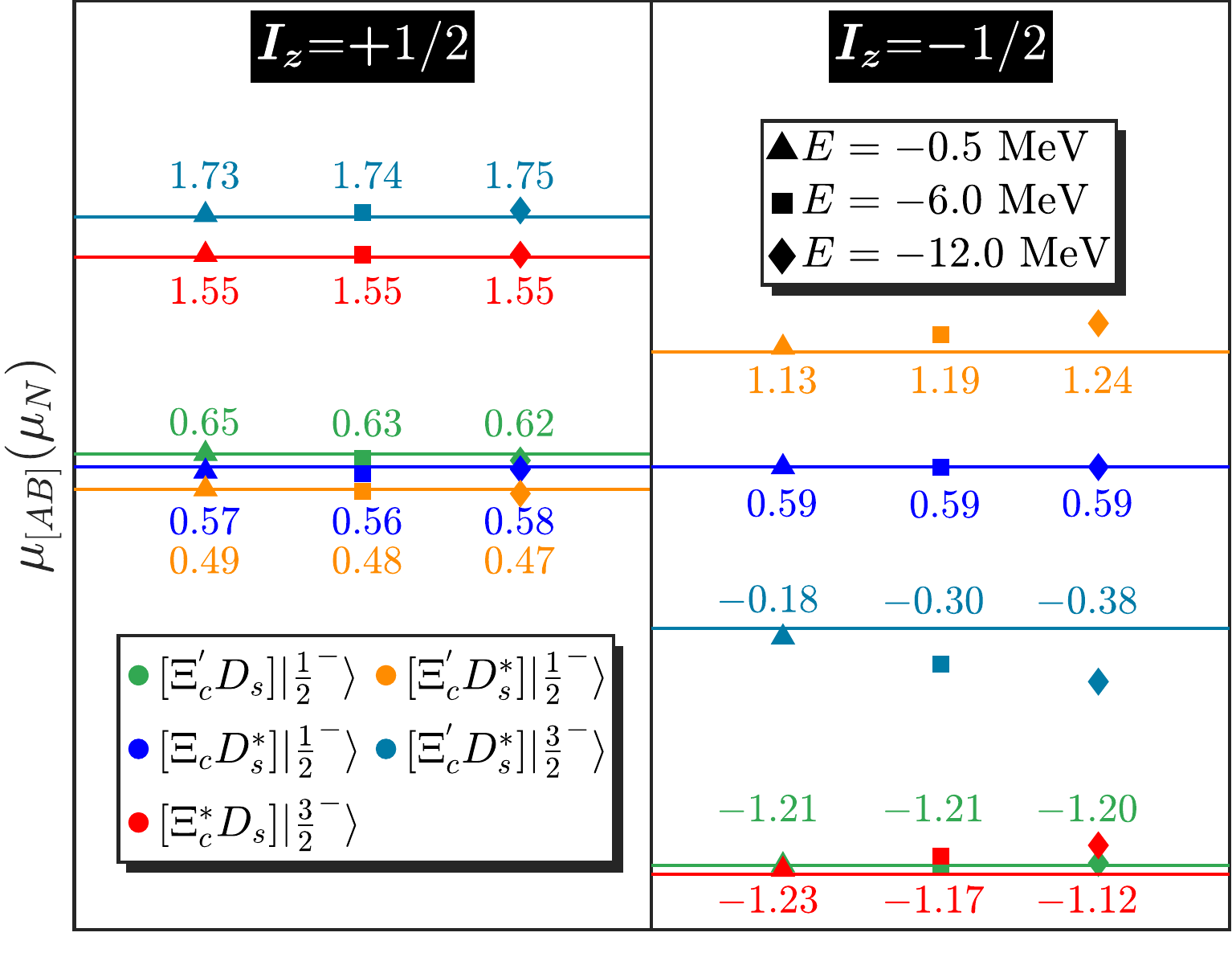}
\caption{The magnetic moments $\mu_{[AB]}$ of the predicted $\Xi_c^{(',*)}D^{(*)}_s$ molecular states in coupled-channel analysis. The corresponding single-channel results are shown as the horizontal lines for comparison.}
\label{fig:couplemu}
\end{figure}

As in the case of $S$-$D$ wave mixing analysis, a rigorous study of the magnetic moments of the  $\Xi_c^{(',*)}D^{(*)}_s$ molecular pentaquarks under coupled-channel effects requires the spatial wave functions of all coupled channels as essential inputs. In the subsequent analysis, we continue to employ three representative binding energies of $-0.5$, $-6.0$, and $-12.0$ MeV to systematically evaluate the corresponding magnetic moments. The results for the magnetic moments of the predicted $\Xi_c^{(',*)}D^{(*)}_s$ molecular candidates in coupled-channel analysis are summarized in Fig. \ref{fig:couplemu}.

A systematic comparison of the magnetic moments of the $\Xi_c^{(',*)}D^{(*)}_s$ molecular pentaquarks obtained from single-channel and coupled-channel analyses reveals that coupled-channel effects can alter their numerical values. In particular, while the magnetic moments of the $\Xi_c D_s^{*}$ state with $J^P = {1/2}^-$ are identical for $I_z = +1/2$ and $I_z = -1/2$ in single-channel analysis, they become distinct once coupled-channel effects are included. {This arises primarily because some channels within the coupled system may exhibit different magnetic moments depending on $I_z$. In addition, the stability of the isospin-dependent effects in the electromagnetic observables of the molecular states observed in the coupled-channel calculations is governed by two key factors: (i) the probability percentages of each channel within the coupled system, which are determined by the mass spectrum of the coupled system, and (ii) the magnetic moments and the transition magnetic moments of the individual channels, which depend on the flavor and spin structures of the constituent hadrons.}

The magnetic moments of the predicted $\Xi_c^{(',*)}D^{(*)}_s$ molecular pentaquark states serve as a sensitive probe of their inner structures and electromagnetic characteristics. Future experimental measurements of these quantities could not only help determine fundamental quantum numbers, such as the spin-parity, the isospin, and so on, but also offer a means to identify the specific constituent hadrons forming the corresponding molecular state.

\section{The M1 radiative decay behavior of the predicted $\Xi_c^{(',*)}D^{(*)}_s$ molecules}\label{section:3}

Building upon our investigation of the magnetic moments of the predicted $\Xi_c^{(',*)}D^{(*)}_s$ molecular pentaquarks, we now turn to their M1 radiative decay properties. As a fundamental electromagnetic observable, the M1 radiative decays probe the transition amplitudes between the hadronic states and offer critical insight into the inner wave function configurations. In this section, we systematically analyze the M1 radiative decay behavior of the $\Xi_c^{(',*)}D^{(*)}_s$ molecular states within the constituent quark model using three complementary approaches: single-channel analysis, $S$-$D$ wave mixing analysis, and coupled-channel analysis.

\subsection{Formalism for calculating the M1 radiative decay width between the hadrons within the constituent quark model}

In the following, we outline the formalism for calculating the M1 radiative decay width between the hadronic molecules within the constituent quark model. The M1 radiative decay width between the hadronic molecules for the ${[AB]} \rightarrow {[AB]}' \gamma$ process, denoted as $\Gamma_{{[AB]} \rightarrow {[AB]}' \gamma}$, is given by the expression \cite{Li:2021ryu,Zhou:2022gra,Wang:2022tib,Wang:2023bek,Lai:2024jfe,Zhang:2025ame,Dey:1994qi,Majethiya:2009vx,Majethiya:2011ry}:
 \begin{eqnarray}\label{eq:transitionmagneticmoment}
     \Gamma_{{[AB]} \rightarrow {[AB]}' \gamma} &=&\frac{\alpha_{\text{EM}}}{2J_{[AB]} + 1} \frac{k^3}{m_p^2}\frac{\sum _{J_{{[AB]}'z},J_{{[AB]}z}}\left(\begin{array}{ccc}
      		J_{{[AB]}'} & 1 & J_{{[AB]}} \\
      		-J_{{[AB]}'z} & 0 & J_{{[AB]}z} \\
      	\end{array}\right)^2}{\left(\begin{array}{ccc}
      		J_{{[AB]}'} & 1 & J_{{[AB]}} \\
      		-J_{z} & 0 & J_{z} \\
      	\end{array}\right)^2}\nonumber\\
    &&\times\frac{|\mu_{{[AB]} \rightarrow {[AB]}' \gamma}|^2}{\mu_N^2}.
\end{eqnarray}
Here, $J_z$ is the smaller of the total angular momentum quantum numbers $J_{[AB]}$ and $J_{{[AB]}'}$ of the initial and final molecules, and $k$ refers to the emitted photon's momentum, given by $k=\frac{m_{[AB]}^2-m^2_{{[AB]}'}}{2m_{[AB]}}$ with $m_{[AB]}$ and $m_{{[AB]}'}$ being the masses of the initial and final molecules, respectively. The proton mass $m_p$ is taken as 938.00 MeV \cite{ParticleDataGroup:2024cfk}, and the electromagnetic fine structure constant $\alpha_{\text{EM}}$ is approximate 1/137. The quantities $\left(\begin{array}{ccc}
      		J_{{[AB]}'} & 1 & J_{{[AB]}} \\
      		-J_{{[AB]}'z} & 0 & J_{{[AB]}z} \\
      	\end{array}\right)$ and $\left(\begin{array}{ccc}
      		J_{{[AB]}'} & 1 & J_{{[AB]}} \\
      		-J_{z} & 0 & J_{z} \\
      	\end{array}\right)$ denote the Wigner 3-$j$ coefficient.

As indicated in Eq.~(\ref{eq:transitionmagneticmoment}), the central quantity in the calculation of the M1 radiative decay width between the hadronic states is the corresponding transition magnetic moment, denoted as $\mu_{{[AB]} \to {[AB]}^{\prime} \gamma}$. In this work, we calculate the transition magnetic moment between the $\Xi_c^{(',*)}D^{(*)}_s$ molecular states using the following expression \cite{Li:2021ryu,Zhou:2022gra,Wang:2022tib,Wang:2023bek,Lai:2024jfe,Zhang:2025ame}:
\begin{eqnarray}
\mu_{{[AB]} \to {[AB]}^{\prime}\gamma}=\left\langle{J_{{[AB]}^{\prime}},J_{z}\left|\sum_{j}\hat{\mu}_{zj}^{\rm spin}e^{-\text{i} {\bf k}\cdot{\bf r}_j}+\hat{\mu}_z^{\rm orbital}\right|J_{{[AB]}},J_{z}}\right\rangle,\nonumber\\
\end{eqnarray}
where the exponential factor $e^{-\text{i}\mathbf{k}\cdot \mathbf{r}_j}$ is the spatial wave function of the emitted photon, $\mathbf{k}$ is the momentum of the emitted photon, and $\mathbf{r}_j$ denotes the coordinate of the $j$-th quark.

In calculating the transition magnetic moment between the $\Xi_c^{(',*)}D^{(*)}_s$ molecular states, an accurate treatment of the spatial wave functions both for the molecular states and their constituent hadrons are essential. Following the approach used in our earlier magnetic moment calculations for these molecular states, we employ their numerical spatial wave functions obtained by solving the coupled-channel Schr\"{o}dinger equation. For the baryons $\Xi_c^{(',*)}$ and the mesons $D^{(*)}_s$, their spatial wave functions are modeled using the simple harmonic oscillator (SHO) form:
\begin{equation}\begin{aligned}
      	\phi_{n,l,m}(\beta,\mathbf{r})= &\sqrt{\frac{2n!}{\Gamma(n+l+\frac{3}{2})}} L_n^{l+\frac{1}{2}}(\beta^2 r^2) \beta^{l+\frac{3}{2}}e^{-\frac{\beta^2 r^2}{2}} r^l Y_{lm}(\Omega),
 \end{aligned}\end{equation}
where $Y_{lm}(\Omega)$ is the spherical harmonic function, $L_n^{l+\frac{1}{2}}(x)$ is the associated Laguerre polynomial, and $n$, $l$, and $m$ are the radial, orbital, and magnetic quantum numbers of the discussed hadrons, respectively. By fitting their experimental mass spectra \cite{ParticleDataGroup:2024cfk}, we determine the oscillator parameters $\beta$ as \cite{Luo:2023sne}:
\begin{eqnarray*}
\left.\begin{array}{c}
(\beta_\rho,\,\beta_\lambda)_{\Xi_c}=(0.301,\,0.383)~{\rm GeV}, \\
(\beta_\rho,\,\beta_\lambda)_{\Xi_c^{'}}=(0.252,\,0.383)~{\rm GeV}, \\
(\beta_\rho,\,\beta_\lambda)_{\Xi_c^*}=(0.243,\,0.358)~{\rm GeV}, \\
\beta_{D_s}=0.429~{\rm GeV},~~~\beta_{D_s^{*}}=0.371~{\rm GeV}.
\end{array}\right.
\end{eqnarray*}
In addition, the spatial wave function of the emitted photon $e^{-\text{i}{\bf k}\cdot{\bf r}}$ can be expanded as \cite{Khersonskii:1988krb}
\begin{eqnarray}
e^{-\text{i}{\bf k}\cdot{\bf r}}=\sum\limits_{l=1}^\infty\sum\limits_{m=-l+1}^{l-1}4\pi(-\text{i})^{l-1}j_{l-1}(kr)Y_{l-1m}^*(\Omega_{\bf k})Y_{l-1m}(\Omega_{{\bf r}}),
\end{eqnarray}
where $j_l(x)$ denotes the spherical Bessel function. This expansion provides a systematic framework for evaluating the spatial overlap matrix element between the initial and final states $\langle\phi_{f}|e^{-\text{i}\mathbf{k}\cdot \mathbf{r}_j}|\phi_{i}\rangle$.

\subsection{Numerical results of the M1 radiative decay widths of the predicted $\Xi_c^{(',*)}D^{(*)}_s$ molecules}

Building upon the established theoretical framework, we now systematically investigate the M1 radiative decay widths of the $\Xi_c^{(',*)}D^{(*)}_s$ molecular pentaquarks. In line with our earlier analysis of their magnetic moments, the study is organized into three distinct approaches: single-channel analysis, $S$-$D$ wave mixing analysis, and coupled-channel analysis.

To establish a reference for discussing the M1 radiative decay widths of the $\Xi_c^{(',*)}D^{(*)}_s$ molecular pentaquarks in relation to their constituent hadrons, we first evaluate the M1 radiative decay widths between the constituent hadrons. These results provide a baseline for comparative analysis. The calculated decay widths are as follows:
\begin{eqnarray*}
\begin{aligned}
&\Gamma_{\Xi_c^{'+} \rightarrow \Xi_c^+\gamma}=\textbf{20.90}\,{\rm keV},~~~~~~~~~&&\Gamma_{\Xi_c^{'0} \rightarrow \Xi_c^0\gamma}=0.16\,{\rm keV},\\
&\Gamma_{\Xi_c^{*+} \rightarrow \Xi_c^+\gamma}=\textbf{75.74}\,{\rm keV},~~~~~~~~~&&\Gamma_{\Xi_c^{*0} \rightarrow \Xi_c^0\gamma}=0.54\,{\rm keV},\\
&\Gamma_{\Xi_c^{*+} \rightarrow \Xi_c^{'+}\gamma}=0.04\,{\rm keV},~~~~~~~~~&&\Gamma_{\Xi_c^{*0} \rightarrow \Xi_c^{'0}\gamma}=1.50\,{\rm keV},\\
&\Gamma_{D_s^{*+} \rightarrow D_s^{+}\gamma}=0.70\,{\rm keV}. ~~~~~~~~~
\end{aligned}
\end{eqnarray*}
Notably, the certain M1 radiative decay processes between the constituent hadrons forming the $\Xi_c^{(',*)}D^{(*)}_s$ molecular states exhibit considerable decay widths, as exemplified by processes $\Xi_c^{*+} \rightarrow \Xi_c^+\gamma$ and $\Xi_c^{'+} \rightarrow \Xi_c^+\gamma$. This suggests that analogous M1 radiative decay channels within the $\Xi_c^{(',*)}D^{(*)}_s$ molecular pentaquarks may also be significant, rendering them promising targets for future experimental detection.

We now turn to the M1 radiative decay widths of the $\Xi_c^{(',*)}D^{(*)}_s$ molecular states in single-channel approximation. In these calculations, the same binding energy is adopted for both the initial and final molecular states in a given M1 radiative transition. To systematically examine the behavior of their M1 decay widths, we employ three representative binding energies: $-0.5$, $-6.0$, and $-12.0$ MeV.

\renewcommand\tabcolsep{0.05cm}
\renewcommand{\arraystretch}{1.50}
\begin{table*}[!htbp]
\centering
\caption{The M1 radiative decay widths $\Gamma_{{[AB]} \rightarrow {[AB]}' \gamma}$ of the predicted $\Xi_c^{(',*)}D^{(*)}_s$ molecular states in single-channel analysis, given in keV. Results are shown for three representative binding energies: $-0.5$, $-6.0$, and $-12.0$ MeV. Processes that are strictly forbidden are marked with \ding{55}, while those with widths below 0.005 keV are indicated as $\mathcal{O}(0)$.}\label{muLsingle}
\begin{tabular}{ccccccccc}
  \toprule[1pt]\toprule[1pt]
$\Gamma_{{[AB]} \rightarrow {[AB]}' \gamma}$ & $I_z$ & $[\Xi_cD_s ]| \frac{1}{2}^- \rangle\gamma$&$[\Xi_c^{'}D_s ]| \frac{1}{2}^- \rangle\gamma$ & $[\Xi_cD_s^{*} ]| \frac{1}{2}^- \rangle\gamma$&$[\Xi_cD_s^{*} ]| \frac{3}{2}^- \rangle\gamma$&$[\Xi_c^{*}D_s ]| \frac{3}{2}^- \rangle\gamma$ & $[\Xi_c^{'}D_s^{*} ]| \frac{1}{2}^- \rangle\gamma$ & $[\Xi_c^{'}D_s^{*} ]| \frac{3}{2}^- \rangle\gamma$ \\ \hline
  \multirow{2}{*}{$[\Xi_c^{'} D_s ]| \frac{1}{2}^- \rangle $} &$+\frac{1}{2}$&\textbf{15.44},\,\textbf{20.52},\,\textbf{20.90}&&&&&&
  \\&$-\frac{1}{2}$&\hphantom{0}0.12,\,\hphantom{0}0.16,\,\hphantom{0}0.16&&&&&&\\
  \multirow{2}{*}{$[\Xi_c D_s^{*} ]| \frac{1}{2}^- \rangle $} &$+\frac{1}{2}$&\hphantom{0}0.39,\,\hphantom{0}0.67,\,\hphantom{0}0.70&\ding{55}&&$\mathcal{O}(0)$,$\,\mathcal{O}(0)$,$\,\mathcal{O}(0)$&&&
  \\&$-\frac{1}{2}$&\hphantom{0}0.39,\,\hphantom{0}0.67,\,\hphantom{0}0.70&\ding{55}&&$\mathcal{O}(0)$,$\,\mathcal{O}(0)$,$\,\mathcal{O}(0)$&&&\\
    \multirow{2}{*}{$[\Xi_c D_s^{*} ]| \frac{3}{2}^- \rangle $} &$+\frac{1}{2}$&\hphantom{0}0.39,\,\hphantom{0}0.67,\,\hphantom{0}0.70&\ding{55}&$\mathcal{O}(0)$,$\,\mathcal{O}(0)$,$\,\mathcal{O}(0)$&&&&
  \\&$-\frac{1}{2}$&\hphantom{0}0.39,\,\hphantom{0}0.67,\,\hphantom{0}0.70&\ding{55}&$\mathcal{O}(0)$,$\,\mathcal{O}(0)$,$\,\mathcal{O}(0)$&&&&\\
    \multirow{2}{*}{$[\Xi_c^* D_s ]| \frac{3}{2}^- \rangle $} &$+\frac{1}{2}$&\textbf{39.05},\,\textbf{70.61},\,\textbf{74.00} &0.04,\,0.04,\,0.04&\ding{55}&\ding{55}&&&
  \\&$-\frac{1}{2}$&\hphantom{0}0.28,\,\hphantom{0}0.50,\,\hphantom{0}0.52 &1.33,\,1.49,\,1.50&\ding{55}&\ding{55}&&&\\
   \multirow{2}{*}{$[\Xi_c^{'} D_s^{*} ]| \frac{1}{2}^- \rangle $} &$+\frac{1}{2}$&\ding{55}&0.38,\,0.64,\,0.66&\hphantom{0}1.69,\,\hphantom{0}2.22,\,\hphantom{0}2.24&\textbf{13.56},\,\textbf{17.79},\,\textbf{17.91}&\ding{55}&&$\mathcal{O}(0)$,$\,\mathcal{O}(0)$,$\,\mathcal{O}(0)$
  \\&$-\frac{1}{2}$&\ding{55}&0.38,\,0.64,\,0.66&\hphantom{0}0.01,\,\hphantom{0}0.02,\,\hphantom{0}0.02&\hphantom{0}0.10,\,\hphantom{0}0.13,\,\hphantom{0}0.14&\ding{55}&&$\mathcal{O}(0)$,$\,\mathcal{O}(0)$,$\,\mathcal{O}(0)$\\
     \multirow{2}{*}{$[\Xi_c^{'} D_s^{*} ]| \frac{3}{2}^- \rangle $} &$+\frac{1}{2}$&\ding{55}&0.40,\,0.67,\,0.70&\hphantom{0}6.93,\,\hphantom{0}9.10,\,\hphantom{0}9.26&\hphantom{0}8.67,\,\textbf{11.38},\,\textbf{11.57}&\ding{55}&$\mathcal{O}(0)$,$\,\mathcal{O}(0)$,$\,\mathcal{O}(0)$&
  \\&$-\frac{1}{2}$&\ding{55}&0.40,\,0.67,\,0.70&\hphantom{0}0.05,\,\hphantom{0}0.07,\,\hphantom{0}0.07&\hphantom{0}0.07,\,\hphantom{0}0.09,\,\hphantom{0}0.09&\ding{55}&$\mathcal{O}(0)$,$\,\mathcal{O}(0)$,$\,\mathcal{O}(0)$&\\
  \multirow{2}{*}{$[\Xi_c^{*} D_s^{*} ]| \frac{1}{2}^- \rangle $} &$+\frac{1}{2}$&\ding{55}&\ding{55}&\textbf{34.19},\,\textbf{60.55},\,\textbf{62.59} &\hphantom{0}4.27,\,\hphantom{0}7.57,\,\hphantom{0}7.82&0.38,\,0.64,\,0.66&0.03,\,0.04,\,0.04&$\mathcal{O}(0)$,$\,\mathcal{O}(0)$,$\,\mathcal{O}(0)$
  \\&$-\frac{1}{2}$&\ding{55}&\ding{55}&\hphantom{0}0.24,\,\hphantom{0}0.43,\,\hphantom{0}0.44 &\hphantom{0}0.03,\,\hphantom{0}0.05,\,\hphantom{0}0.06 &0.38,\,0.64,\,0.66&1.18,\,1.33,\,1.34&0.15,\,0.16,\,0.15\\
     \multirow{2}{*}{$[\Xi_c^{*} D_s^{*} ]| \frac{3}{2}^- \rangle $} &$+\frac{1}{2}$&\ding{55}&\ding{55}&\textbf{21.65},\,\textbf{38.84},\,\textbf{40.50}&\textbf{17.32},\,\textbf{31.07},\,\textbf{32.40}&0.39,\,0.66,\,0.68&0.02,\,0.02,\,0.02 &0.02,\,0.02,\,0.02
  \\&$-\frac{1}{2}$&\ding{55}&\ding{55}&\hphantom{0}0.15,\,\hphantom{0}0.28,\,\hphantom{0}0.29&\hphantom{0}0.12,\,\hphantom{0}0.22,\,\hphantom{0}0.23 &0.39,\,0.66,\,0.68&0.74,\,0.83,\,0.83&0.59,\,0.65,\,0.64\\
\multirow{2}{*}{$[\Xi_c^{*} D_s^{*} ]| \frac{5}{2}^- \rangle $} &$+\frac{1}{2}$&\ding{55}&\ding{55}&\ding{55}&\textbf{37.88},\,\textbf{70.65},\,\textbf{73.75} &0.38,\,0.67,\,0.69 &\ding{55}&0.04,\,0.04,\,0.04
\\&$-\frac{1}{2}$&\ding{55}&\ding{55}&\ding{55}&\hphantom{0}0.27,\,\hphantom{0}0.50,\,\hphantom{0}0.52&0.38,\,0.67,\,0.69 &\ding{55}&1.33,\,1.50,\,1.50\\
     \toprule[1pt]\toprule[1pt]
    \end{tabular}
\end{table*}

In Table \ref{muLsingle}, we summarize the M1 radiative decay widths of the predicted $\Xi_c^{(',*)}D^{(*)}_s$ molecular candidates for single-channel analysis.  It is noteworthy that
\begin{itemize}
  \item Among the M1 radiative decay processes of the $\Xi_c^{(',*)}D^{(*)}_s$ molecular states, several processes exhibit notably large decay widths. Representative examples with $I_z = 1/2$ include:  $[\Xi_c^{'} D_s] |{1}/{2}^- \rangle \to [\Xi_cD_s ]|{1}/{2}^- \rangle\gamma$, $[\Xi_c^{*} D_s] |{3}/{2}^- \rangle \to [\Xi_cD_s ]|{1}/{2}^- \rangle\gamma$, $[\Xi_c^{*} D_s^{*}]| {1}/{2}^- \rangle \to [\Xi_cD_s^{*} ]|{1}/{2}^- \rangle\gamma$, $[\Xi_c^{*} D_s^{*}] |{3}/{2}^- \rangle \to [\Xi_cD_s^{*} ]|{1}/{2}^- \rangle\gamma$,
  $[\Xi_c^{'} D_s^{*}] |{1}/{2}^- \rangle \to [\Xi_cD_s^{*} ]|{3}/{2}^- \rangle\gamma$, $[\Xi_c^{'} D_s^{*}] |{3}/{2}^- \rangle \to [\Xi_cD_s^{*} ]|{3}/{2}^- \rangle\gamma$, $[\Xi_c^{*} D_s^{*}] |{3}/{2}^- \rangle \to [\Xi_cD_s^{*} ]|{3}/{2}^- \rangle\gamma$, and $[\Xi_c^{*} D_s^{*}]|{5}/{2}^- \rangle \to [\Xi_cD_s^{*} ]|{3}/{2}^- \rangle\gamma$, and so on. This pattern can be largely attributed to the sizable M1 radiative decay widths of the constituent hadrons within these discussed molecular states. The observation implies that the M1 decay behavior of the molecular states effectively reflects the electromagnetic properties of its components, offering a useful guiding principle for predicting such decay modes in the molecular states.
  \item The M1 radiative decay widths of the $\Xi_c^{(',*)}D^{(*)}_s$ molecular states depend sensitively on both the transition magnetic moments and the available kinematic phase space. For example, the processes $[\Xi_{c}^{'}D_s^{*}]|1/2^{-}\rangle \to [\Xi_{c}D_s^{*}]|1/2^{-}\rangle \gamma$ and $[\Xi_{c}^{*}D_s^{*}]|1/2^{-}\rangle \to [\Xi_{c}D_s^{*}]|3/2^{-}\rangle \gamma$ have the transition magnetic moments of similar magnitude. However, the former exhibits a significantly smaller decay width, which can be explained by its more limited phase space. On the other hand, the decays $[\Xi_{c}^{*}D_s^{*}]|1/2^{-}\rangle \to [\Xi_{c}D_s^{*}]|1/2^{-}\rangle \gamma$ and $[\Xi_{c}^{*}D_s^{*}]|1/2^{-}\rangle \to [\Xi_{c}D_s^{*}]|3/2^{-}\rangle \gamma$ occur under identical phase space conditions, yet the former displays a markedly larger M1 radiative decay width, a direct consequence of its stronger transition magnetic moment.
  \item M1 radiative decays can serve as a sensitive probe for determining both the spin-parity quantum numbers and the underlying constituent hadrons of the $\Xi_c^{(',*)}D^{(*)}_s$ molecular states in future experiments. For instance, the decays $[\Xi_c^{*} D_s^{*} ]|{1}/{2}^- \rangle \to [\Xi_cD_s^{*}]|{3}/{2}^- \rangle\gamma$, $[\Xi_c^{*} D_s^{*} ]|{3}/{2}^- \rangle \to [\Xi_cD_s^{*}]|{3}/{2}^- \rangle\gamma$, and $[\Xi_c^{*} D_s^{*} ]|{5}/{2}^- \rangle \to [\Xi_cD_s^{*}]|{3}/{2}^- \rangle\gamma$ exhibit markedly different M1 radiative decay widths. This distinct pattern enables the use of the M1 radiative behavior to disentangle the spin-parity assignments of the $\Xi_c^{*} D_s^{*}$ molecular states. Moreover, a pronounced contrast is observed between the M1 radiative decay widths of the processes $[\Xi_c^{\prime} D_s^{*} ]|{1}/{2}^- \rangle \to [\Xi_cD_s^{*}]|{1}/{2}^- \rangle\gamma$ and $[\Xi_c^{*} D_s^{*} ]|{1}/{2}^- \rangle \to [\Xi_cD_s^{*}]|{1}/{2}^- \rangle\gamma$. This clear disparity offers an unambiguous spectroscopic signature, thereby providing a mean to discriminate between different constituent hadronic configurations through analysis of their M1 radiative decay properties.
\end{itemize}

{Here, it is important to note that the oscillator parameters $\beta$ characterize the spatial wave functions of the baryons $\Xi_c^{(',*)}$ and the mesons $D^{(*)}_s$ serve as key inputs in evaluating the electromagnetic properties of the $\Xi_c^{(',*)}D^{(*)}_s$ molecular states. In this work, the $\beta$ values for each hadrons are determined by fitting their experimental mass spectra. Given the uncertainties associated with these parameters, it is necessary to assess their impact on the electromagnetic properties of the $\Xi_c^{(',*)}D^{(*)}_s$ molecular states. Following a similar approach to the treatment of the constituent quark masses, we examine two typical radiative decay processes by varying $\beta$ within $\pm 10\%$ in single-channel analysis. The numerical results are as follows:
\begin{eqnarray*}
\Gamma_{[\Xi_c^{'} D_s] |{1}/{2}^- \rangle \to [\Xi_cD_s ]|{1}/{2}^- \rangle\gamma}^{I_z = 1/2}&=&20.52^{+1.05}_{-4.08}{~\rm keV},\\
\Gamma_{[\Xi_c^{*} D_s] |{3}/{2}^- \rangle \to [\Xi_cD_s ]|{1}/{2}^- \rangle\gamma}^{I_z = 1/2}&=&70.61^{+5.74}_{-17.15}{~\rm keV}.
\end{eqnarray*}
In the calculations, the binding energies of both the initial and final molecular states are taken to be
$-6.0$ MeV. The resulting uncertainties in the radiative decay widths arising from the variation of
$\beta$ are indicated as the superscripts and subscripts attached to the central values. While the numerical values of the radiative decay widths may shift to some extent when the oscillator parameters are varied, the order of magnitude generally remains stable. This point has been discussed in more detail in our previous work \cite{Wang:2022nqs}.}

We now investigate the M1 radiative decay behavior of the predicted $\Xi_c^{(',*)}D^{(*)}_s$ molecular states by including contributions from the $D$-wave channels. A comprehensive set of the corresponding decay widths of the predicted $\Xi_c^{(',*)}D^{(*)}_s$ molecular candidates obtained in $S$-$D$ wave mixing analysis is provided in Table \ref{muLsd}.

Compared to single-channel results, the inclusion of $S$-$D$ wave mixing effects has a negligible impact on the M1 radiative decay widths of the $\Xi_c^{(',*)}D^{(*)}_s$ molecular candidates. This result aligns with our earlier analysis of their magnetic moments and reflects very small $D$-wave admixture in the total spatial wave function, which in turn leads to only minor changes in the electromagnetic observables.

\renewcommand\tabcolsep{0.35cm}
\renewcommand{\arraystretch}{1.50}
\begin{table}[!htbp]
\centering
\caption{The M1 radiative decay widths $\Gamma_{{[AB]} \rightarrow {[AB]}' \gamma}$ of the predicted $\Xi_c^{(',*)}D^{(*)}_s$ molecular states obtained in $S$-$D$ wave mixing analysis. All values are given in keV and are evaluated at three representative binding energies:  $-0.5$, $-6.0$, and $-12.0$ MeV. Processes that are strictly forbidden are marked with \ding{55}, while those with widths below 0.005 keV are labeled as $\mathcal{O}(0)$.}\label{muLsd}
\begin{tabular}{cccc}
  \toprule[1pt]\toprule[1pt]
$\Gamma_{{[AB]} \rightarrow {[AB]}' \gamma}$ & $I_z$ & $[\Xi_c^{'}D_s^{*} ]| \frac{1}{2}^- \rangle\gamma$ & $[\Xi_c^{'}D_s^{*} ]| \frac{3}{2}^- \rangle\gamma$ \\ [1pt] \hline
  \multirow{2}{*}{$[\Xi_c^{*} D_s^{*} ]| \frac{1}{2}^- \rangle $} &$+\frac{1}{2}$&0.03,\,0.04,\,0.04& $\mathcal{O}(0)$,$\,\mathcal{O}(0)$,$\,\mathcal{O}(0)$\\
  &$-\frac{1}{2}$&1.18,\,1.33,\,1.34&0.14,\,0.16,\,0.15\\
   \multirow{2}{*}{$[\Xi_c^{*} D_s^{*} ]| \frac{3}{2}^- \rangle $} &$+\frac{1}{2}$ &0.02,\,0.02,\,0.02&0.02,$\,$0.02,$\,$0.02\\
    &$-\frac{1}{2}$&0.74,\,0.82,\,0.82&0.58,\,0.64,\,0.64\\
   \multirow{2}{*}{$[\Xi_c^{*} D_s^{*} ]| \frac{5}{2}^- \rangle $ }&$+\frac{1}{2}$ & \ding{55} &0.04,\,0.04,\,0.04\\
    &$-\frac{1}{2}$&\ding{55}&1.31,\,1.50,\,1.50\\
     \toprule[1pt]\toprule[1pt]
    \end{tabular}
\end{table}

Having analyzed the M1 radiative decay behavior of the $\Xi_c^{(',*)}D^{(*)}_s$ molecular pentaquarks through both single-channel analysis and $S$-$D$ wave mixing analysis, we proceed to include coupled-channel effects. The corresponding decay widths are summarized in Table~\ref{muLcouple}.

\renewcommand\tabcolsep{0.40cm}
\renewcommand{\arraystretch}{1.50}
\begin{table*}[!htbp]
\centering
\caption{The M1 radiative decay widths $\Gamma_{{[AB]} \rightarrow {[AB]}' \gamma}$ of the predicted $\Xi_c^{(',*)}D^{(*)}_s$ molecular states obtained in coupled-channel analysis. All values are given in keV and correspond to three representative binding energies: $-0.5$, $-6.0$, and $-12.0$ MeV. Processes that are strictly forbidden are marked with \ding{55}, while those with widths below 0.005 keV are denoted as $\mathcal{O}(0)$.}\label{muLcouple}
\begin{tabular}{ccccccc}
  \toprule[1pt]\toprule[1pt]
 $\Gamma_{{[AB]} \rightarrow {[AB]}' \gamma}$& $I_z$ & $[\Xi_c^{'}D_s ]| \frac{1}{2}^- \rangle\gamma$ & $[\Xi_cD_s^{*} ]| \frac{1}{2}^- \rangle\gamma$&$[\Xi_c^{*}D_s ]| \frac{3}{2}^- \rangle\gamma$ & $[\Xi_c^{'}D_s^{*} ]| \frac{1}{2}^- \rangle\gamma$ & $[\Xi_c^{'}D_s^{*} ]| \frac{3}{2}^- \rangle\gamma$ \\ [1pt] \hline
  \multirow{2}{*}{$[\Xi_c D_s^{*} ]| \frac{1}{2}^- \rangle $} &$+\frac{1}{2}$&$\mathcal{O}(0)$,\,0.01,\,0.01 &&&&
  \\&$-\frac{1}{2}$&$\mathcal{O}(0)$,$\,\mathcal{O}(0)$,$\,\mathcal{O}(0)$&&&&\\
  \multirow{2}{*}{$[\Xi_c^{*} D_s ]| \frac{3}{2}^- \rangle $} &$+\frac{1}{2}$&0.04,\,0.03,\,0.02&$\mathcal{O}(0)$,$\,\mathcal{O}(0)$,$\,\mathcal{O}(0)$&&&
  \\&$-\frac{1}{2}$&1.35,\,1.56,\,1.61&$\mathcal{O}(0)$,$\,\mathcal{O}(0)$,$\,\mathcal{O}(0)$&&&\\
  \multirow{2}{*}{$[\Xi_c^{'} D_s^{*} ]| \frac{1}{2}^- \rangle $} &$+\frac{1}{2}$&0.30,\,0.29,\,0.22&\hphantom{0}1.57,\,\hphantom{0}1.48,\,\hphantom{0}1.03&$\mathcal{O}(0)$,$\,\mathcal{O}(0)$,\,0.01 &&$\mathcal{O}(0)$,$\,\mathcal{O}(0)$,$\,\mathcal{O}(0)$
  \\&$-\frac{1}{2}$&0.33,\,0.28,\,0.13&\hphantom{0}0.02,\,\hphantom{0}0.02,\,\hphantom{0}0.01&0.01,\,0.11,\,0.29&&$\mathcal{O}(0)$,$\,\mathcal{O}(0)$,$\,\mathcal{O}(0)$\\
  \multirow{2}{*}{$[\Xi_c^{'} D_s^{*} ]| \frac{3}{2}^- \rangle $} &$+\frac{1}{2}$&0.54,\,0.99,\,1.00 &\hphantom{0}7.32,\,\textbf{11.10},\,\textbf{12.24}&$\mathcal{O}(0)$,\,0.04,\,0.10 &$\mathcal{O}(0)$,$\,\mathcal{O}(0)$,$\,\mathcal{O}(0)$&
  \\&$-\frac{1}{2}$&0.60,\,1.78,\,2.59&\hphantom{0}0.07,\,\hphantom{0}0.11,\,\hphantom{0}0.10&$\mathcal{O}(0)$,\,0.01,\,0.03 &$\mathcal{O}(0)$,$\,\mathcal{O}(0)$,$\,\mathcal{O}(0)$&\\
  \multirow{2}{*}{$[\Xi_c^{*} D_s^{*} ]| \frac{1}{2}^- \rangle $} &$+\frac{1}{2}$&0.05,\,0.34,\,0.49&\textbf{33.84},\,\textbf{60.07},\,\textbf{62.04}&0.36,\,0.56,\,0.51&0.03,\,0.02,\,0.01&$\mathcal{O}(0)$,$\,\mathcal{O}(0)$,$\,\mathcal{O}(0)$
  \\&$-\frac{1}{2}$&0.03,\,0.49,\,1.12&\hphantom{0}0.20,\,\hphantom{0}0.31,\,\hphantom{0}0.34&0.28,\,0.18,\,0.05&1.15,\,1.11,\,0.94&0.18,\,0.33,\,0.44 \\
  \multirow{2}{*}{$[\Xi_c^{*} D_s^{*} ]| \frac{3}{2}^- \rangle $} &$+\frac{1}{2}$&0.03,\,0.16,\,0.20&\textbf{21.48},\,\textbf{38.72},\,\textbf{40.33}&0.27,\,0.16,\,0.04&0.02,\,0.02,\,0.02&0.03,\,0.09,\,0.15
  \\&$-\frac{1}{2}$&$\mathcal{O}(0)$,$\,\mathcal{O}(0)$,$\,\mathcal{O}(0)$&\hphantom{0}0.15,\,\hphantom{0}0.27,\,\hphantom{0}0.28&0.36,\,0.51,\,0.43&0.76,\,1.00,\,1.16&0.61,\,0.72,\,0.75\\
  \multirow{2}{*}{$[\Xi_c^{*} D_s^{*} ]| \frac{5}{2}^- \rangle $} &$+\frac{1}{2}$&\ding{55}&\ding{55}&0.36,\,0.58,\,0.55 &\ding{55}&0.04,\,0.04,\,0.04
  \\&$-\frac{1}{2}$&\ding{55}&\ding{55}&0.61,\,2.42,\,3.94&\ding{55}&1.23,\,1.14,\,0.98 \\
     \toprule[1pt]\toprule[1pt]
    \end{tabular}
\end{table*}

A comparative analysis of the M1 radiative decay behavior in the $\Xi_c^{(',*)}D^{(*)}_s$ molecular pentaquarks reveals that coupled-channel effects can influence these processes when comparing results from single-channel analysis and $S$-$D$ wave mixing analysis. Notably, coupled-channel effects play a particularly non-trivial role in specific M1 radiative decay processes of the $\Xi_c^{(',*)}D^{(*)}_s$ molecular states. This particularly manifests in transitions where the interference between different hadronic configurations modifies the decay widths. For instance, while single-channel analysis yields the same M1 radiative decay widths for the $[\Xi_c^{'} D_s^{*}]|1/2^{-}\rangle \to [\Xi_c^{'}D_s ]|1/2^{-}\rangle\gamma$ process with $I_z=+1/2$ and $I_z=-1/2$, coupled-channel analysis reveals pronounced differences. An analogous behavior is observed in the processes $[\Xi_c^{'} D_s^{*}]|3/2^{-}\rangle \to [\Xi_c^{'}D_s ]|1/2^{-}\rangle\gamma$,
$[\Xi_c^{*} D_s^{*}]|1/2^{-}\rangle \to [\Xi_c^{*}D_s ]|3/2^{-}\rangle\gamma$, $[\Xi_c^{*} D_s^{*}]|3/2^{-}\rangle \to [\Xi_c^{*}D_s ]|3/2^{-}\rangle\gamma$, and $[\Xi_c^{*} D_s^{*}]|5/2^{-}\rangle \to [\Xi_c^{*}D_s ]|3/2^{-}\rangle\gamma$.

{To provide further guidance for experimental searches, we present the following discussion. Analogous to the well-established double-charm tetraquark state $T_{cc}(3875)^+$ \cite{LHCb:2021auc,LHCb:2021vvq}, we suggest that the predicted $\Xi_c^{(',*)}D^{(*)}_s$ molecular pentaquarks can be produced in the $pp$ collisions at the LHCb. As argued in Refs. \cite{Ozdem:2025olj,Ozdem:2026wmf}, the hadrons with relatively large magnetic moments are expected to couple more strongly to external electromagnetic fields, thereby enhancing their production in the photon-induced channels. The magnetic moments of the order of 1 $\mu_N$ obtained for the $\Xi_c^{(',*)}D^{(*)}_s$ molecular states indicate that the electromagnetic production mechanisms may be within experimental reach, for instance via $\gamma p \to P_{ccs \bar s} X$ at facilities such as GlueX or future electron-ion colliders. A more direct probe of the electromagnetic structures of the $P_{ccs \bar s}$ states lies in their radiative decays. These can be observed in processes such as $e^+ e^- \to P_{ccs \bar s}\gamma X$ or $pp \to P_{ccs \bar s}\gamma X$, at experiments like Belle II and LHCb, with the radiative decay widths serving as the key observables. To enable a full reconstruction chain, the decays of the $\Xi_c^{(',*)}D^{(*)}_s$ molecular states can be traced through the sequential decays of their constituents. The constituent baryons may be reconstructed using the main decay channels, for example, $\Xi_c^{*} \to \Xi_c \pi$, $\Xi_c^{\prime} \to \Xi_c \gamma$, $\Xi_c \to \Xi \pi$, followed by $\Xi \to \Lambda \pi$ and $\Lambda \to p \pi$. The constituent mesons can be identified via decays such as $D_s^*\to D_s \gamma \to \phi(1020)\pi \gamma\to K K \pi \gamma$.}

\section{summary} \label{section:4}

Since the discovery of the hidden-charm molecular pentaquarks and the double-charm molecular tetraquark, the study of the mass spectra of the double-charm molecular pentaquarks has emerged as a major research focus in the field of hadron physics. Previous theoretical work \cite{Yalikun:2023waw} has predicted the mass spectra of the $\Xi_c^{(',*)}D^{(*)}_s$-type double-charm hidden-strangeness molecular pentaquark candidates with $J^P={1/2}^-$, ${3/2}^-$, and ${5/2}^-$ within the OBE model through the incorporation of both $S$-$D$ wave mixing effects and coupled-channel effects. In this work, we conduct a systematic and comprehensive investigation into of the electromagnetic properties of the predicted $\Xi_c^{(',*)}D^{(*)}_s$ molecular pentaquarks within the constituent quark model, focusing specifically on their magnetic moments and M1 radiative decay behavior.

As fundamental electromagnetic property of hadrons, we first discuss the magnetic moments of the predicted $\Xi_c^{(',*)}D^{(*)}_s$ molecular pentaquarks. In single-channel case, we find that the magnetic moments of the $\Xi_c^{(',*)}D^{(*)}_s$ molecular states are given by the sum of the magnetic moments of their constituent hadrons and can be used to discriminate both the spin-parity quantum numbers and the constituent hadronic configurations. The introduction of $S$-$D$ wave mixing effects was found to have a negligible influence, as the minimal proportion of the $D$-wave component in the total spatial wave function. In contrast, we find that coupled-channel effects play a non-trivial role in specific states, such as the magnetic moment can distinguish the $z$-component of the isospin of the $\Xi_c D^{*}_s$ molecular state with $J^P={1/2}^-$.  Our results indicate that the magnetic moments of the predicted $\Xi_c^{(',*)}D^{(*)}_s$ molecules serve as sensitive probes of their inner structures, exhibiting patterns that can discriminate between different constituent configurations and spin-parity quantum number assignments.

And then, we explore the M1 radiative decay behavior of the predicted $\Xi_c^{(',*)}D^{(*)}_s$ molecular pentaquarks along the same analytical hierarchy. The investigation of their M1 radiative decay widths has yielded equally important insights. In single-channel analysis, several transitions exhibit considerable M1 radiative decay widths that may serve as promising signatures for experimental detection, which are largely governed by the intrinsic transitions of the constituent hardons. The processes with different spin-parity transition and same constituent hadrons transition or with same spin-parity transition and different constituent hadrons tansition, which exhibit significant difference of their M1 radiative decay widths, can used to discriminate the spin-parity quantum numbers and the constituent hadronic configurations of the $\Xi_c^{(',*)}D^{(*)}_s$ molecular states. Similar to the conclusions of the magnetic moments, $S$-$D$ wave mixing effects are minimal. However, coupled-channel effect can influence their M1 radiative decay widths, particularly in the $[\Xi_c^{'} D_s^{*}] \to [\Xi_c^{'}D_s ]\gamma$ and $[\Xi_c^{*} D_s^{*}] \to [\Xi_c^{*}D_s ]\gamma$ processes with $I_z=\pm 1/2$.

In conclusion, the electromagnetic properties of the predicted $\Xi_c^{(',*)}D^{(*)}_s$ molecular states provide distinctive signatures that complement information obtained from the mass spectra \cite{Yalikun:2023waw}, offering important criteria for testing their molecular nature. Our systematic study of the electromagnetic properties of the $\Xi_c^{(',*)}D^{(*)}_s$ molecular states has not only provided specific predictions for experimental verification but also demonstrated the power of the electromagnetic probes in elucidating the structures and dynamics of these $\Xi_c^{(',*)}D^{(*)}_s$-type double-charm hidden-strangeness molecular pentaquark candidates. Therefore, we propose that the future experiments should focus on investigating the electromagnetic properties of the hadronic molecular states, as such studies hold crucial implications for understanding their inner structures and fundamental characteristics.

{In this work, we focus specifically on the electromagnetic properties of the $\Xi_c^{(',*)}D^{(*)}_s$ molecular states. Our results demonstrate that these observables are highly sensitive to the spin-parity assignments of the hadronic molecules, thus offering valuable theoretical guidance for future experimental determination of their spin-parity quantum numbers. It would be particularly valuable to investigate the electromagnetic properties of the compact double-charm hidden-strangeness pentaquarks in future studies, as such work could help clarify the inner configurations of the double-charm hidden-strangeness pentaquarks.
Currently, the experimental study of the electromagnetic properties of the hadrons, particularly for the exotic states, faces significant challenges. Nevertheless, recent progress in lattice QCD calculations has begun to address the electromagnetic properties of the exotic states \cite{Vujmilovic:2025czt}, providing valuable theoretical benchmarks and motivating further experimental and theoretical investigations. It is worth noting that the strong decay widths of these molecular states might be significantly larger than their radiative decay widths. Looking ahead, we strongly encourage further studies employing a variety of models and methods to explore other properties of these hadronic molecules, including their strong decay behaviors. A comprehensive understanding of their inner structures and eventual experimental identification will ultimately rely on the combination of different types of observables, each offering complementary insights.}

\section*{ACKNOWLEDGMENTS}

This work is supported by the National Natural Science Foundation of China under Grant Nos. 12335001, 12247101, and 12405097, the `111 Center' under Grant No. B20063, the Natural Science Foundation of Gansu Province (No. 22JR5RA389, No. 25JRRA799), the Talent Scientific Fund of Lanzhou University, the fundamental Research Funds for the Central Universities (No. lzujbky-2023-stlt01), the project for top-notch innovative talents of Gansu province, and Lanzhou City High-Level Talent Funding.

\end{document}